 \documentclass[preprint,5p,times,twocolumn]{elsarticle}

\usepackage{amssymb}
\usepackage{lipsum}
\usepackage{graphicx}
\usepackage{longtable}
\usepackage{float}

\usepackage[export]{adjustbox}
\usepackage{xurl}
\usepackage{enumitem}
\usepackage{amsmath}
\usepackage{tabularx}
\usepackage[skip=0.5ex]{subcaption}
\usepackage{tfrupee}
\usepackage{float}
\usepackage[title]{appendix}
\usepackage[misc,geometry]{ifsym}

\usepackage{xcolor}
\usepackage{hyperref}

\journal{ArXiv}
\begin{document}
\begin{frontmatter}



\title{Longitudinal Volumetric Study for the Progression of Alzheimer's Disease from Structural MRI}


\author[label1]{Prayas Sanyal\corref{corr1}} \ead{prayassanyal008@gmail.com} 

\author[label1]{Srinjay Mukherjee} \author[label1]{Arkapravo Das} \author[label1]{Anindya Sen\fnref{label2}} \ead{anindya.sen@heritageit.edu}
\affiliation[label1]{organization={Dept. of Electronics and Communication Engineering, Heritage Institute of Technology},  
            city={Kolkata},
            country={India}}
\cortext[corr1]{Corresponding Author} \fntext[label2]{Senior Author}

\begin{abstract}
Alzheimer's Disease (AD) is an irreversible neurodegenerative disorder affecting millions of individuals today. The prognosis of the disease solely depends on treating symptoms as they arise and proper caregiving, as there are no current medical preventative treatments apart from newly developing drugs which can, at most, slow the progression. Thus, early detection of the disease at its most premature state is of paramount importance. This work aims to survey imaging biomarkers corresponding to the progression of AD and also reviews some of the existing feature extraction methods. A longitudinal study of structural MR images was performed for given temporal test subjects with AD selected randomly from the Alzheimer's Disease Neuroimaging Initiative (ADNI) database. A pipeline was implemented to study the data, including modern pre-processing techniques such as spatial image registration, skull stripping, inhomogeneity correction and tissue segmentation using an unsupervised learning approach using intensity histogram information. The temporal data across multiple visits is used to study the structural change in volumes of these tissue classes, namely, cerebrospinal fluid (CSF), grey matter (GM), and white matter (WM) as the patients progressed further into the disease. To detect changes in volume trends, we also analyse the data with a modified Mann-Kendall statistic. The segmented features thus extracted and the subsequent trend analysis provide insights such as atrophy, increase or intolerable shifting of GM, WM and CSF and should help in future research for automated analysis of Alzheimer's detection with clinical domain explainability.
\end{abstract}




\end{frontmatter}




\section{Introduction}
\label{introduction}
Reports show that 8.8 million Indians above the age of 60 live with dementia (Alzheimer's Disease is a more specific subtype of dementia), amounting to approximately 7.4\% of the total population. It is projected that the number of dementia afflicted may increase to 16.9 million by 2036 with the rapid rise of correlated risk factors such as diabetes and hypertension, both under-treated in rural settings~\cite{LeeJ2023,Ravi2021}. Common AD symptoms include memory impairment, loss of cognitive function, depression and paranoia. Even though these symptoms inevitably worsen over time, those providing care feel more in control with an early diagnosis and seek assistance from support groups along with others who have experienced similar misfortunes~\cite{Anil2022}. A review published in 2013 states that the annual overhead cost of looking after the elderly with Alzheimer's in India can reach heights of \rupee2,02,450 in urban areas and \rupee66,025 in rural areas, including medication cost, consultation and hospitalization and intangible costs such as loss of productivity and trauma.~\cite{GRao2013}.

\begin{figure}[!h]
  \centering
 \begin{tabular}{ccc}
    \hspace{-1.7cm}
    \begin{subfigure}{0.4\textwidth}
      \centering
      \includegraphics[width=0.49\linewidth,height = 3.1cm]{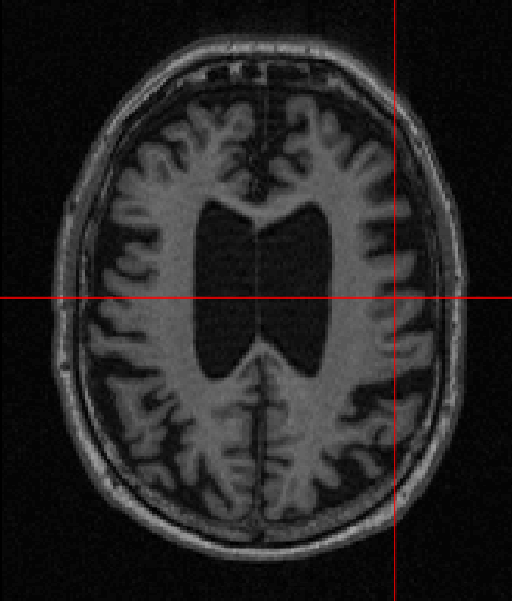}
      \caption{Axial plane}
    \end{subfigure}
    \hspace{-3cm}
    \begin{subfigure}{0.4\textwidth}
      \centering
      \includegraphics[width=0.49\linewidth,height = 3.1cm]{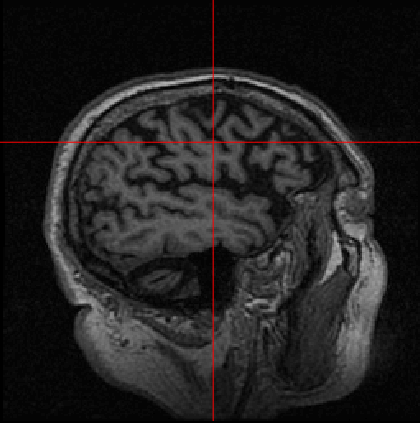}
      \caption{Sagittal plane}
    \end{subfigure}  \\
    \hspace{-1.7cm}
    \begin{subfigure}{0.4\textwidth}
      \centering
      \includegraphics[width=0.49\linewidth,height = 3.1cm]{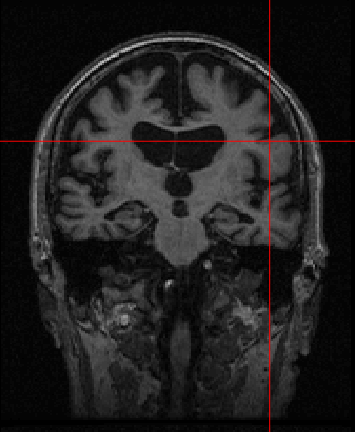}
      \caption{Coronal plane}
    \end{subfigure} 
    \end{tabular}
  \captionsetup{justification=centering}
  \caption{\small{MP-RAGE T1 Weighted MR slices}}
  \label{fig:MRIEg}
\end{figure}

German psychiatrist and neuropathologist Alois Alzheimer laid the foundations of most of the modern understanding of the underlying causes of this disorder when he noticed the common biomarkers through histological techniques on the brains of his diseased patients~\cite{AloisOriginal1907}.\newline Fig.~\ref{fig:MRIEg} shows the example of an AD patient's raw unprocessed brain slice in different planes obtained from MR imaging in a magnetization-prepared rapid gradient-echo (MP-RAGE) sequence. Marking the cerebral atrophy and other imaging biomarkers over time helps in the timely prognosis of Alzheimer's. 
High-resolution activation of brain regions during various cognitive tasks is also often used to understand the progression of this particular disease. As discussed in the following section, Magnetic Resonance Imaging (MRI) is one of the frequently used modalities to detect brain atrophy, whereas functional MRI may be used to check various patterns of brain activations~\cite{Ogawa1990}. In this paper, we have focused on structural MRI scans only. 
\subsection{Existing Works}
Multiple large-scale longitudinal studies have been carried out to investigate the causal connection between various AD indicators and the development of cognitive decline and dementia (e.g., ADNI, Australian Imaging, MIRIAD)~\cite{BondiSurvey2017,JackCR2013,MIRIAD}. Due to the dearth of publicly available datasets corresponding only to the Indian populace, the ADNI database was extensively used for our experimentation. Over the past two decades, several techniques have been applied to measuring brain atrophy in structural MRI, such as the boundary shift integral technique~\cite{BShift1,BShift2}, which uses a linear registration to align base and follow-up scans to track the change in brain boundary; or Jacobian integration~\cite{Jacobian} as an improvement to BSI technique, or the more recent tensor-based morphometry (TBM)~\cite{TBM1} which utilises non-linear image registration to map the volume change between the baseline and follow up images. Numerous machine-learning feature extraction techniques have also been studied for distinctive biomarker detection in the advent of AD, but most lack reproducibility and generalizability in the current scenario~\cite{ML2020,ML2015,ML2014}.\newline \textbf{The primary objective of the paper is summarised as follows:} We review current existing methods of extracting tissue features from structural Magnetic Resonance Imaging (MRI) data and propose a user-friendly yet explainable wrapper pipeline in R software for volume extraction in the context of Alzheimer's Disease. This adds an extra step in volume-based feature extraction to aid in future AI automated analysis, which cannot explain certain trends in observed data for clinicians. We consider a publicly available (post a written proposal acceptance) dataset called ADNI to experiment and evaluate our approach for a limited amount of disease affected patients due to computational limitations, which can be extended for future works to gather more insights. Our interpretable pipeline provides clinicians with trend analysis for patients suffering from Alzheimer's. Given the high annotation costs of tissue labels and limited time constraints for clinical experts, this pipeline will help to explain statistical patterns in the quest for understanding the progression of Alzheimer's Disease.

\section{Methodology}
\label{Methodology}
\begin{figure}[!htb]
    \includegraphics[width = 0.49\textwidth,height = 12cm]{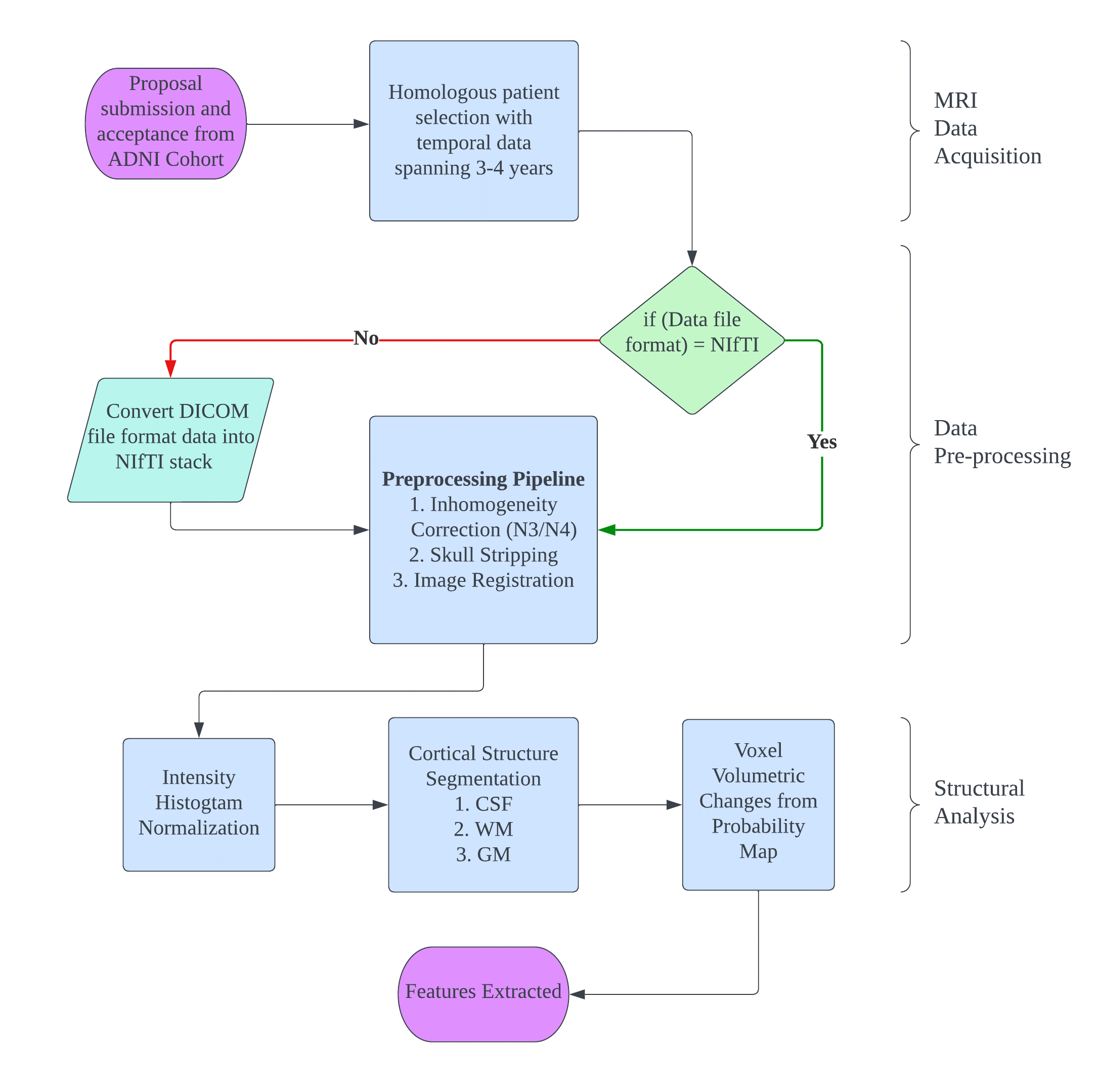}
    \caption{{Flowchart for Structural Analysis in MP-RAGE MR images}}
    \label{fig:Flowchart}
\end{figure}

We put forward a simple script to extract imaging features from MP-RAGE MRI images related to Alzheimer's Disease, which include medical domain explainability for future group pattern analysis works. For each patient, a standard pipeline was implemented in the open-source R software (integrated with fslr~\cite{fslR}). Fig~\ref{fig:Flowchart} represents a flowchart to elucidate the steps such as database formation, pre-processing of the images, and feature extraction. We have only analysed the volumetric changes of three particular tissues, namely, cerebrospinal fluid (CSF), grey matter (GM) and white matter (WM), over the course of multiple visits for a patient diagnosed with AD. These volumetric data thus obtained are considered to be the features of Alzheimer's Disease.

\begin{figure*}[!ht]
  \centering
    \begin{subfigure}{0.39\textwidth}
      \centering
      \includegraphics[width=\linewidth,height = 4.8cm]{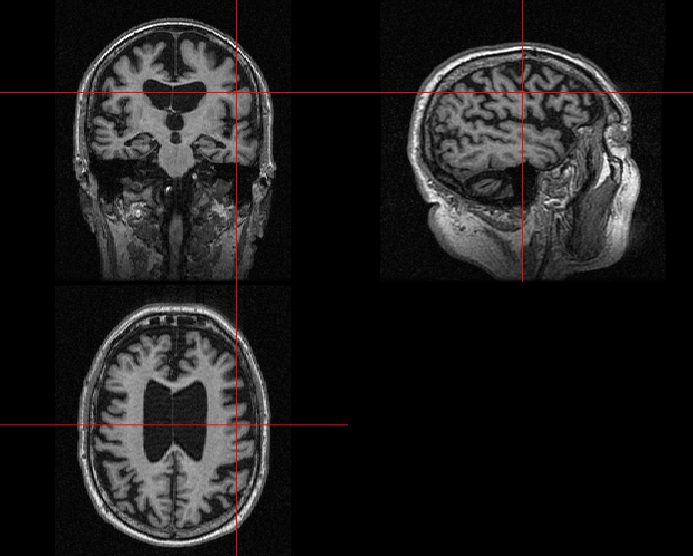}
      \subcaption{Source Scan}
      \label{fig:inhomogenietyA}
    \end{subfigure} 
    \begin{subfigure}{0.39\textwidth}
      \centering
      \includegraphics[width=\linewidth,height = 4.8cm]{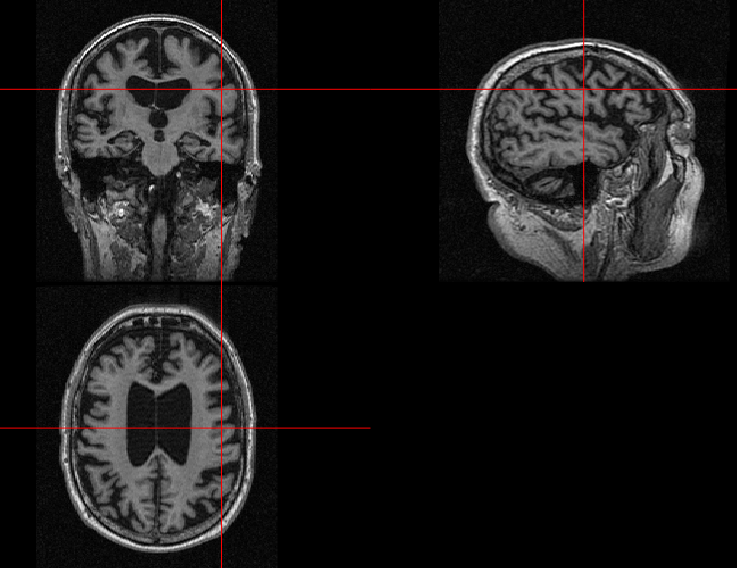}
      \subcaption{Bias Field Corrected Scan}
      \label{fig:inhomogeneityB}
    \end{subfigure} \\
    \begin{subfigure}{0.39\textwidth}
      \centering
      \includegraphics[width=\linewidth,height = 4.8cm]{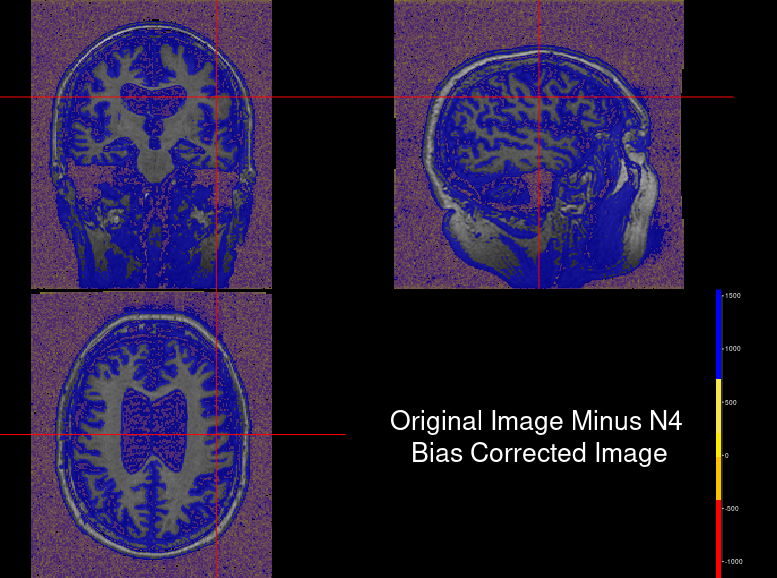}
      \subcaption{Bias field corrected scan subtracted from source scan}
    \label{fig:inhomogeneityC}
    \end{subfigure} 
  \captionsetup{justification=centering}
  \caption{{Inhomogeneity Correction}}
  \label{fig:Inhomogeneity}
\end{figure*}

\begin{figure*}[!ht]
  \centering
  \begin{tabular}{cc}
    \begin{subfigure}{0.38\textwidth}
      \centering
      \includegraphics[width=\linewidth,height = 4.8cm]{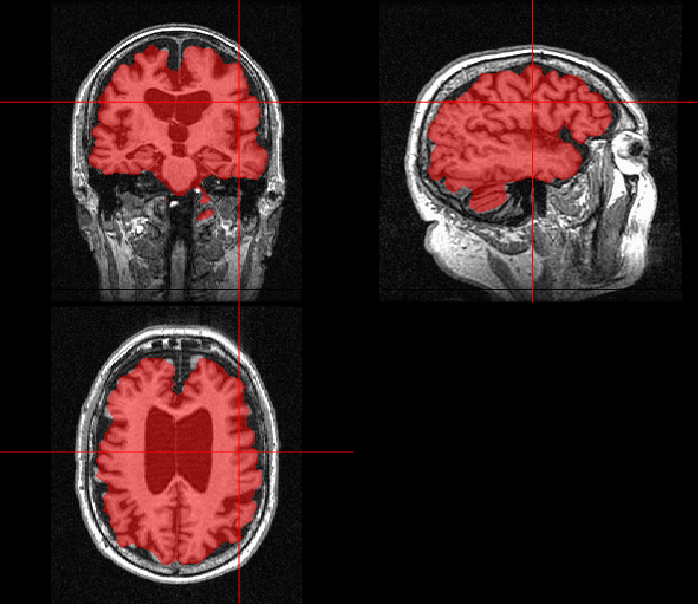}
      \caption{Brain tissue mask overlay}
    \end{subfigure} & 
    \begin{subfigure}{0.38\textwidth}
      \centering
      \includegraphics[width=\linewidth,height = 4.8cm]{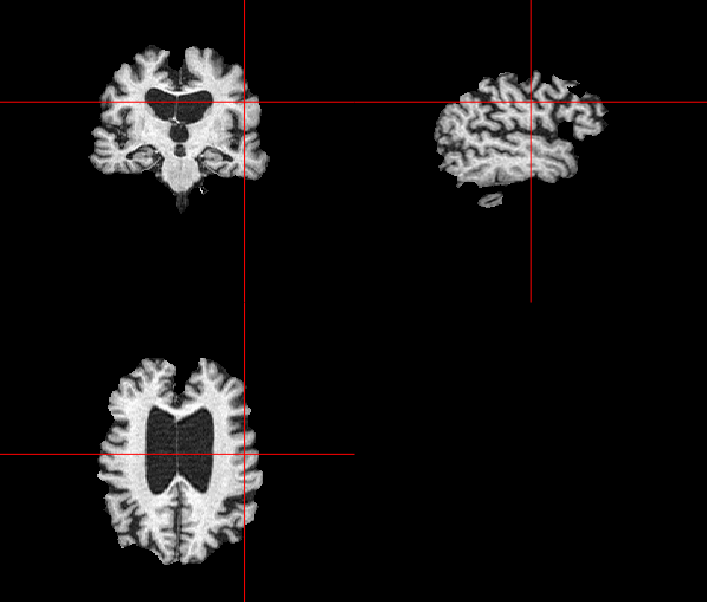}
      \caption{Extracted Brain Tissue}
    \end{subfigure}
  \end{tabular}
  \captionsetup{justification=centering}
  \caption{{Skull Stripping}}
  \label{fig:SkullStrip}
\end{figure*}
\subsection{Database Used}
A representative database was created by selecting T1-weighted MP-RAGE MRI scans of Alzheimer's affected patients at random after gaining access to the Alzheimer's Disease Neuroimaging Initiative (ADNI: \url{adni.loni.usc.edu}), which was launched in 2003 and led by Principal Investigator Michael W. Weiner, MD. The patients were chosen such that they had at least three visits to track the change in the features of their disease progression.

\subsection{Pre-processing}
This section explains in detail the pre-processing paradigm implemented to align all the different visits. A wrapper script thus created with each of the steps enables us to easily process incoming files and perform the collection of transformations to further analyze our data.

\subsubsection{File Preparation}

The primary acquisition plane for each of the scans was sagittal, and the MP-RAGE scans downloaded from the ADNI website were in Digital Imaging and Communications in Medicine (DICOM) format. The DICOM format files (256 separated files for each slice) were first converted into a NIfTI stack (1 single file) to represent each visit for easier processing and visualization. In DICOM image files, the data comprises 2D layers, whereas NIfTI (Neuroimaging Informatics Technology Initiative) is a file format where images and other data are stored in a 3D structure. To access the DICOM files and convert them into a NIfTI file format, the $oro.dicom$ and $oro.nifti$ packages in R are used, which was introduced by Brandon Witcher et al.~\cite{WhitcherR}.

\subsubsection{Inhomogeneity Correction}

A significant issue faced while analyzing structural properties of the brain is the non-uniformity of intensity in structural MR images. Any low-frequency intensity which can be non-uniformly present in the data is referred to as inhomogeneity or bias field. A standard inhomogeneity correction algorithm known as the N4ITK or Improved Bias Correction~\cite{N4ITK} was used to remove low-frequency intensities. The image formation model of N3 is given by the following equation:
\begin{equation}
    v(x) = u(x)f(x) + n(x)
\end{equation} where $v$ is the input image, $u$ is the uncorrupted actual image, $x$ is any location in the image, $f$ is the bias field, and $n$ is noise(which is assumed to be independent and gaussian). 

As the data is log-transformed and assumes a noise-free scenario ($n(x)=0$), the previous equation becomes \begin{equation}
    log(v(x)) = log(u(x)) + log(f(x))
\end{equation} The N4 algorithm then uses a robust B-spline approximation algorithm of the bias field and iterates until a convergence criterion is met where the modified bias field is approximately similar to the last performed iteration. After the updated transformation, the image is outputted back into the original file using inverse log transform. 
To visualise the effects of bias field correction Fig~\ref{fig:Inhomogeneity} shows a sample patient scan in all three planes by applying N4 bias correction using ANTsR library in R.

Even though visually, there might not be a stark difference between the original images and the corrected scan, white and grey matter intensities are uniform in distribution compared to the source image. Fig~\ref{fig:inhomogeneityC} helps to visualize the difference where the blue gradient represents higher differences. It becomes imperative how there was a slowly varying field existing in the source image, which was removed in the corrected version.

\subsubsection{Skull Stripping}
This pre-processing technique removes the skull voxels from the T1-weighted MRI scan. To do this operation on the N4 bias-corrected image, FSL software~\cite{FSL} was used, which is integrated into R using the fslR package. Fslr performs required operations on NIfTI image objects using FSL commands and returns them as R objects~\cite{fslR}. The function $fslr::fslbet$ is readily available, which takes in the N4 corrected NIfTI image and calls the FSL BET function to perform skull stripping. Brain Extraction Tool (BET)~\cite{BET} uses a deformable model that evolves to fit into the brain tissue surface locally adaptive model forces. To form a brain tissue mask which can be overlayed into the input scan, a NIfTI mask array is created using dimensions information from the header file. Secondly, the areas of the skull-stripped image containing the brain tissue were determined, or every pixel with a positive intensity value, because the skull stripping algorithm BET sets everything that is not brain to binary zeroes. A centre of gravity (CoG) of the extracted tissue was determined during the first iteration, and the algorithm was repeated by feeding the CoG as additional information to improve the skull stripping paradigm as one single iteration was observed to contain deformities. Bias correction, which is explained in the previous subsection, is essential for this step as the skull stripping algorithm heavily depends on the tissue intensities, and a low variation in the field can hinder the output. The skull-stripped image thus obtained is shown in Fig~\ref{fig:SkullStrip}.

\begin{figure*}[h]
    \centering
    \includegraphics[height = 11cm, width = 0.99\textwidth]{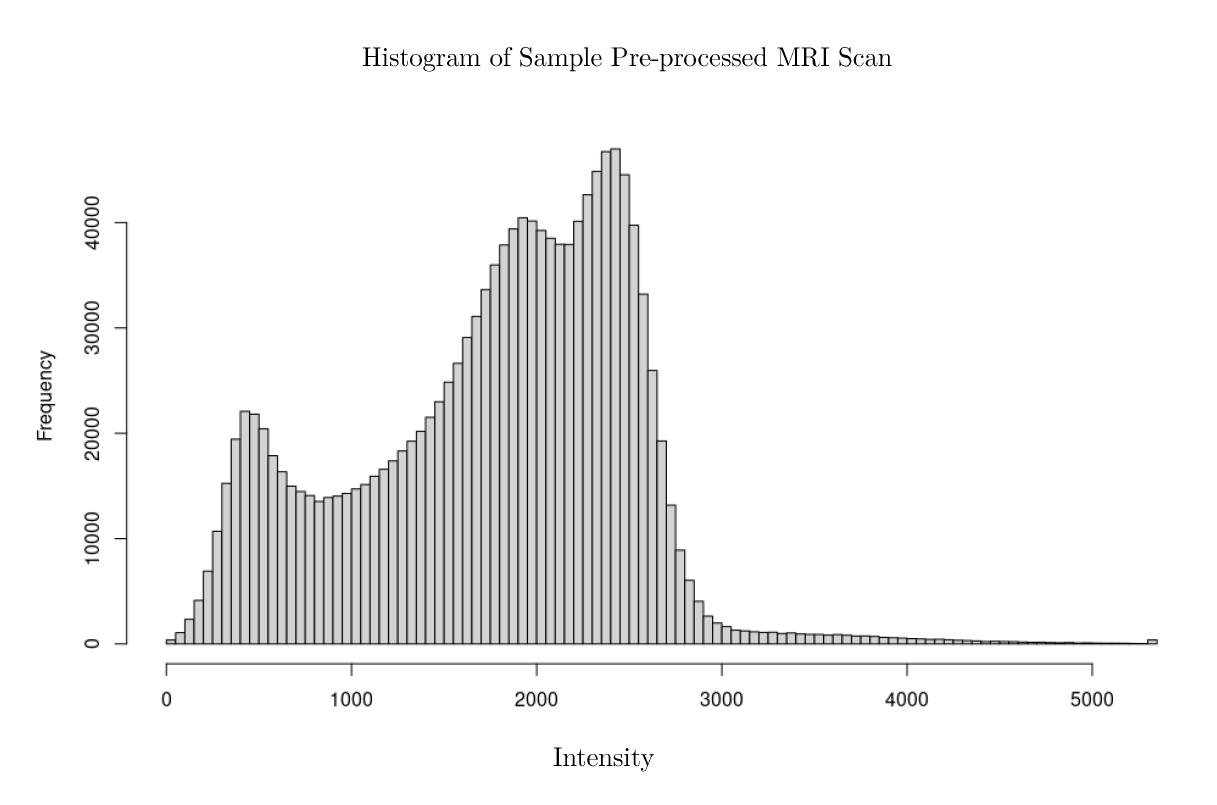}
    \caption{\small{Intensity histogram of a sample AD patient scan after pre-processing}}
    \label{fig:Hist}
\end{figure*}

\subsubsection{Image Registration to Base Scan}
The AD-affected patient scans selected with three or four visits need to be registered to make image locations such as voxels and tissues have similar interpretations and align them spatially to compare changes. Subsequent visit of each patient was co-registered to their source visit after performing inhomogeneity correction and skull stripping. Very few degrees of freedom are needed to register follow-up scans to the base scan as they belong to the same modality, so a rigid or affine registration is sufficient. Rigid registration is a linear registration technique containing six degrees of freedom and is represented by the equation: \begin{equation}
    T_{rigid}(v) = R_v + t
\end{equation}, where t is a translation vector $t = (t_x,t_y,t_z)$ and R is a rotational matrix. All of the above operations were performed using the ANTsR package available in R.

\subsection{Tissue Segmentation and Volume Extraction}
After applying the necessary pre-processing steps described in the former subsections on the MR scans, a three-class tissue segmentation is performed using the FAST algorithm on the FSLR package integrated into R~\cite{fslR,FAST}. The FAST algorithm, which uses the pipeline as described by Zhang et al., puts forward a hidden Markov random field model (HMRF) and an expectation maximization algorithm to perform probabilistic tissue segmentation. The intensity histogram is important for observing the working of the algorithm, so a sample intensity histogram is shown in Fig~\ref{fig:Hist} of an Alzheimer's affected patient's MRI scan after preprocessing (no bias field, noise removed, skull stripped). 
\begin{figure*}[!ht]
  \centering
    \begin{subfigure}{0.4\textwidth}
      \centering
      \includegraphics[width=\linewidth,height = 5.0cm]{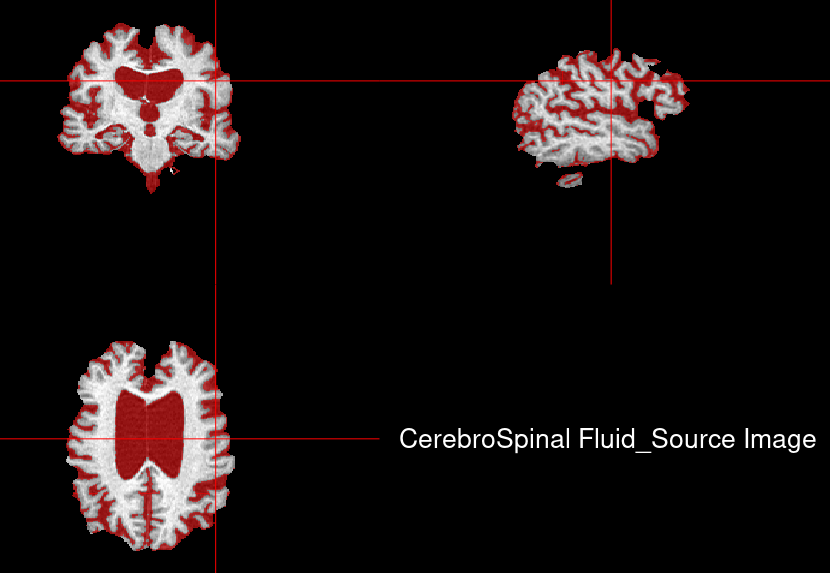}
      \caption{Cerebrospinal Fluid} 
    \end{subfigure} 
    \begin{subfigure}{0.4\textwidth}
      \centering
      \includegraphics[width=\linewidth,height = 5.0cm]{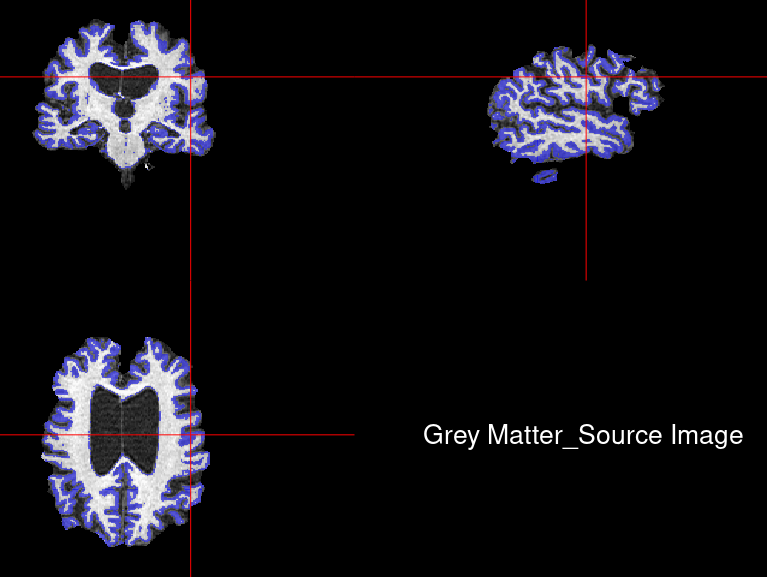}
      \caption{Grey Matter}
    \end{subfigure}
 
    \begin{subfigure}{0.4\textwidth}
      \centering
      \includegraphics[width=\linewidth,height = 5.0cm]{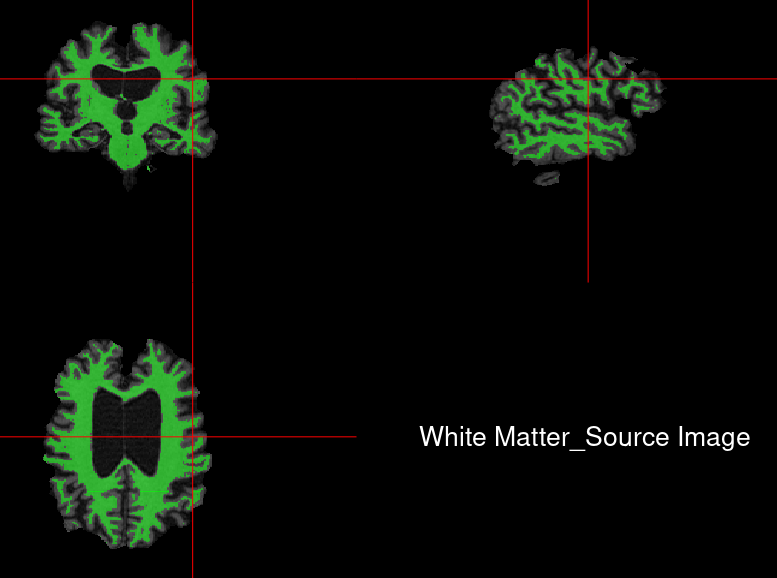}
      \caption{White Matter}
    \end{subfigure} 
  \captionsetup{justification=centering}
  \caption{\small{Segmentation of tissue classes of sample AD patient slice}}
  \label{fig:Segment}
\end{figure*}

If the voxel frequency vs. intensity plot is observed in Fig~\ref{fig:Hist}, the three different tissue classes are apparent as three Gaussian peaks. The lowest intensity peak corresponds to Cerebrospinal Fluid (CSF: $\approx$ 500 unit intensity), whereas the subsequent peaks determine grey matter (GM: $\approx$ 2000 unit intensity) and white matter (WM: $\approx$ 2500 unit intensity). Each of the intensity peaks is widely separated from each other, so there is minimal intra-class variance. To account for spatial neighbourhood information, a hidden Markov Random Field (HMRF) model is used to segment each tissue class as simple classifiers like K-means do not use spatial entropy information, and we need a proper balance. 

The HMRF model as defined in \cite{FAST} is given by: 
\begin{equation}
p\left(y_i \mid x_{\mathcal{N}_i}, \theta\right)=\sum_{\ell \in \mathcal{L}} g\left(y_i ; \theta_{\ell}\right) p\left(\ell \mid X_{\mathcal{N}_i}\right)
\end{equation}
Here, $g(y;\theta_l)$ is a Gaussian probability density function which models the likelihood of $y_i$ (the pixel intensity) given the latent variable $l$ and is given by the equation :
\begin{equation}
g\left(y ; \theta_{\ell}\right)=\frac{1}{\sigma_{\ell}\sqrt{2 \pi }} \exp \left(-\frac{\left(y-\mu_{\ell}\right)^2}{2 \sigma_{\ell}^2}\right) .
\end{equation} 

The parameters associated with the latent variable $l$ are represented by ${\theta}_l = (\mu_l, \sigma_l)^T$. The HMRF model assigns each pixel a certain class label among the three tissues from the set $L$, and each pixel is characterized by an intensity value $y_i$. $\theta$ is a parameter set used as a random variable to compute the marginal probability distribution. The unknown parameters in the equation are estimated using an expectation maximization (EM) algorithm, which is defined in \cite{FAST}. In the expectation (E) step, we initiate each Gaussian cluster with mean $\mu_{\ell}$, Covariance $\sum_{\ell}$ and size of $\pi_{\ell}$. 

For each voxel intensity $y_i$, we compute the probability it belongs to the cluster $l$ and normalise the sum over all the distributions accordingly by:

\begin{equation}
p_{i l}=\frac{\pi_l \mathcal{N}\left(y_i ; \mu_l, \Sigma_l\right)}{\sum_{l^{\prime}} \pi_{l^{\prime}} \mathcal{N}\left(y_i ; \mu_{l^{\prime}}, \Sigma_{l^{\prime}}\right)}
\end{equation}

The parameters ($\mu_{\ell}$, $\sum_{\ell}$,$\pi_{\ell}$) for each cluster are updated using the maximization (M) step. Output files of the FAST algorithm provide three separate probability maps for corresponding tissue classes - $pve_0$: \begin{math}\mathcal{P}\end{math}(CSF = near zero intensity), $pve_1$: \begin{math}\mathcal{P}\end{math}(GM = low to moderate intensity), $pve_2$: \begin{math}\mathcal{P}\end{math}(WM = moderate to high intensity). The $ortho2$ function in R is used to overlay the tissue probability maps on the pre-processed T1 weighted image, as shown in Fig~\ref{fig:Segment}. The changes in volumes of each patient for the tissue classes across three or four visits are observed and reported. These changes can be considered as relevant imaging features to understand the prognosis of Alzheimer's disease.

\begin{figure*}[!ht]
  \centering
    \begin{subfigure}{0.49\textwidth}
      \centering
      \includegraphics[width=1\textwidth,height = 0.67\linewidth]{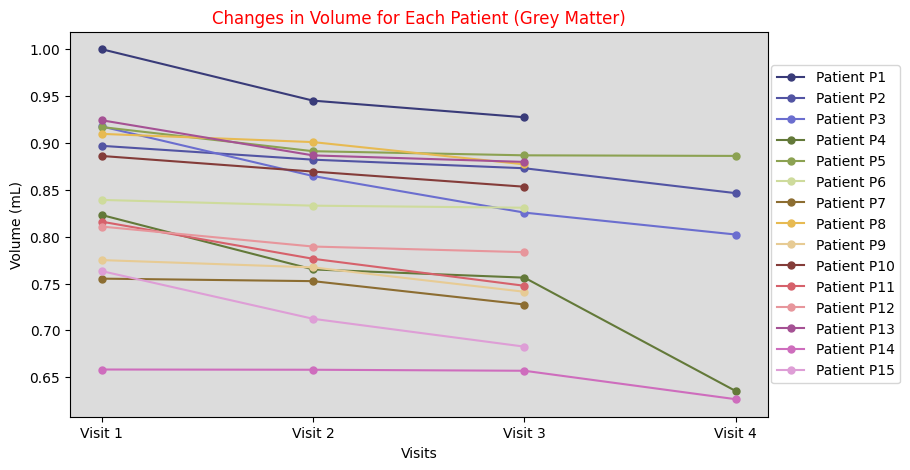}
      \subcaption{Grey matter volume changes}
      \label{fig:gmvolume}
    \end{subfigure}
    \begin{subfigure}{0.49\textwidth}
      \centering
      \includegraphics[width=1\textwidth,height = 0.67\linewidth]{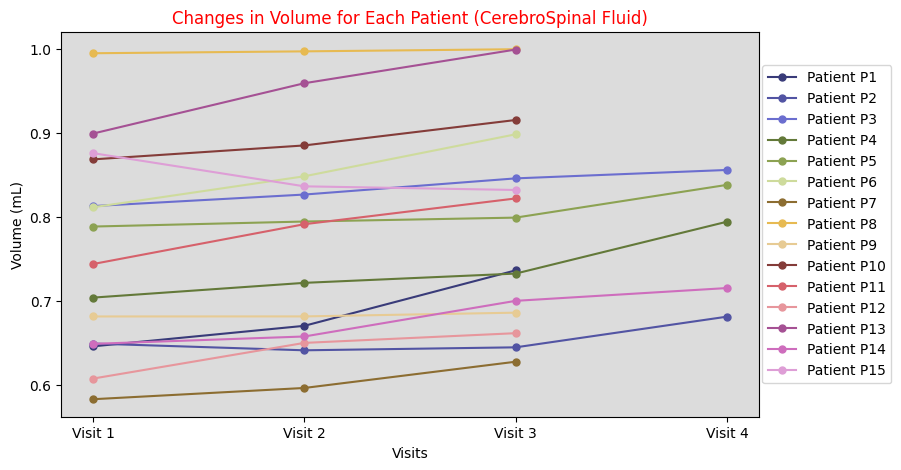}
      \subcaption{Cerebrospinal fluid volume changes}
      \label{fig:csfvolume}
    \end{subfigure} 
     \begin{subfigure}{0.5\textwidth}
      \centering
      \includegraphics[width=1\textwidth,height = 0.67\linewidth]{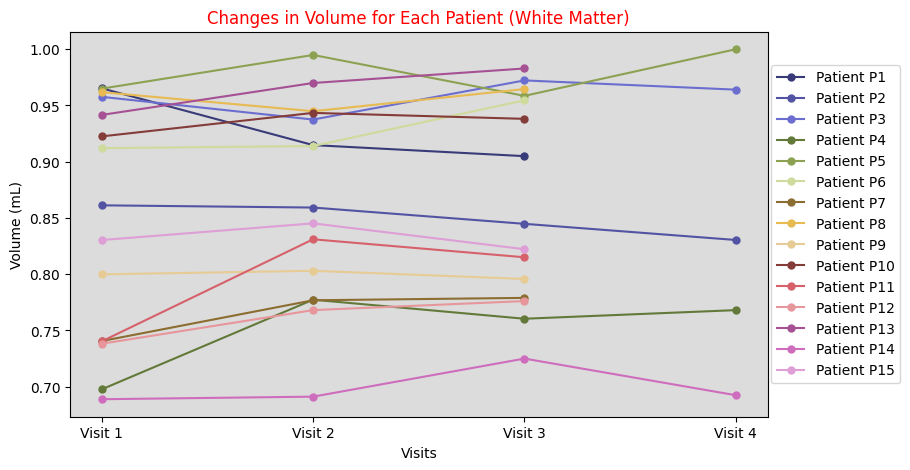}
      \subcaption{White matter volume changes}
      \label{fig:wmvolume}
    \end{subfigure}
      \captionsetup{justification=centering}
  \caption{{Line plot for volume changes of each tissue class}}
  \label{fig:Volume}
  \end{figure*}
\section{Experimental Results}
The above pipeline was iterated for each of the visits of the patients chosen from the ADNI database. To understand, the probability model used in the script can
simply be represented as \begin{equation}
    \log p(intensity) + \beta \log p(HMRF)
\end{equation}, where $\beta$ is user adjustable. $\beta$ was empirically chosen to be 0.4 as the threshold for the output tissue probability maps after multiple experiments to calculate the volumes of each tissue class for the respective scans to extract features. First, the voxel dimension was calculated with a function called $voxdim$ available in $oro.nifti$ package. Then the number of voxels above the chosen threshold is calculated corresponding to the probability maps for each tissue class, given by $t_{vox}$. Finally, the volume is calculated with the equation $Volume = vox_{dim} \times (t_{vox} \div 1000)$, where $vox_{dim}$ is the voxel dimensions in $(mL)$ and the respective tissue voxels are $t_{vox}$ which is normalized to $1000$. 
The changes in volume in each tissue class were plotted in Fig~\ref{fig:Volume} with each line depicting a patient after max scaling the data. Twenty patients' volumetric data were observed, but after several permutations, fifteen of them were chosen to be plotted to show a majority trend (either increase in volume or decrease). 

The mean volume and standard deviation for each subsequent patient visit corresponding to each tissue class were calculated and reported in Table~\ref{tab:gmv}, Table~\ref{tab:csfv}, and Table~\ref{tab:wmv}.

\begin{table}[!h]
    \centering
    \begin{tabular}{|c|c|c|c|c|} \hline 
         Grey Matter&  Visit 1&  Visit 2&  Visit 3& Visit 4\\ \hline 
         Volume (mL)&  554.21&  536.83&  526.15& 497.34\\ \hline 
         Std. Dev.&  57.08&  52.87&  53.54& 79.24\\ \hline
    \end{tabular}
    \caption{Grey Matter Volume (Mean $\pm$ Standard Deviation)}
    \label{tab:gmv}
\end{table}
\vspace{-0.37cm}
\begin{table}[!h]
    \centering
    \begin{tabular}{|c|c|c|c|c|} \hline 
         CSF&  Visit 1&  Visit 2&  Visit 3& Visit 4\\ \hline 
         Volume (mL)&  379.31&  387.44&  398.97& 390.72\\ \hline 
         Std. Dev.&  61.29&  61.58&  61.70& 38.39\\ \hline
    \end{tabular}
    \caption{Cerebrospinal Fluid Volume (Mean$\pm$ Standard Deviation)}
    \label{tab:csfv}
\end{table}
\vspace{-0.37cm}
\begin{table}[!h]
    \centering
    \begin{tabular}{|c|c|c|c|c|} \hline 
         White Matter&  Visit 1&  Visit 2&  Visit 3& Visit 4\\ \hline 
         Volume (mL)&  547.21&  557.89&  558.88& 549.02\\ \hline 
         Std. Dev.&  68.28&  57.47&  58.32& 83.78\\ \hline
    \end{tabular}
    \caption{White Matter Volume (Mean $\pm$ Standard Deviation)}
    \label{tab:wmv}
\end{table}

Since the mean and standard deviation is not a sufficient metric to deduce the change in volumes for individual cases, we have performed a trend analysis to observe a monotonic increase or decrease in tissue volumes in the following section.

\subsection{Trend Analysis for Volumetric Data: Mann-Kendall Test}

To account for increases or decreases in tissue volumes for individual patients, we did the Mann-Kendall (MK) test (Mann, 1945~\cite{Mann} and Kendall, 1976~\cite{Kendall}), which is a non-parametric statistical analysis often used to observe trends in longitudinal time-series data such as ours. Outliers present in the data are not much affected in the overall study as the Mann-Kendall test usually represents monotonic linear dependence~\cite{lazante,sangharsh}. Essentially, the Mann-Kendall test we employ will initially assume the null hypothesis ($H_o$), which is that no detectable monotonic trend is present in the volumetric data, and the alternative hypothesis ($H_u$ or $H_d$) is that there is an upward or downward trend in the respective volumes.
 
\subsubsection{Modified Mann-Kendall Test}

Firstly, we compute the sign of all possible differences by modifying the original MK sign according to our specific needs (Gilbert 1987~\cite{Gilbert}). In a slight modification to the original MK test, a positive one is assigned if the subsequent volume is more than the previous by a magnitude of 1mL instead of all positive volume differences; a negative one is assigned if the subsequent value is less than -1mL in place of all negative volume differences, and a zero is assigned if the difference is between -1mL and 1mL instead of nil-difference changes. This modification is done to ignore minute changes within the 1mL threshold and helps in outlier removal.
We list our volumetric data according to their visits over time $x_1, x_2, . . ., x_n$, where $x_1$ is the volume at visit 1. 
\begin{equation}
\operatorname{sgn}\left(x_j-x_k\right)=\left\{\begin{array}{cl}
+1 & \text { if } x_j-x_k> 1mL \\
0 & \text { if } -1mL <=x_j-x_k<= 1mL \\
-1 & \text { if } x_j-x_k< -1mL
\end{array}\right.
\end{equation}

Now we calculate the Mann Kendall statistic S~\cite{sensSlope} for our new ranked differences between the volumes as per the below equation:

\begin{equation}
S=\sum_{k=1}^{n-1} \sum_{j=k+1}^n \operatorname{sgn}\left(x_j-x_k\right)
\end{equation}

The variance of the MK statistic, denoted by Var(S) is an important variable to find the statistical significance of the monotonic changes. Var(S) is calculated by:

\begin{equation}
\operatorname{Var}(S)=[n(n-1)(2 n+5)]\frac{1}{18}
\end{equation}

To find out the confidence or the statistical significance level corresponding to the cumulative distribution function of the standard normal~\cite{Gilbert}, the $Z_{vol}$ metric value is calculated:

\begin{equation}
Z_{vol}=\left\{\begin{array}{cl}
\frac{S-1}{\sqrt{Var(S)}} \text { if } S>0 \\ \\
0 \quad \text { if } S=0 \\ \\
\frac{S+1}{\sqrt{Var(S)}} \text { if } S<0
\end{array}\right.
\label{zvalue}
\end{equation}

A positive value for $Z_{vol}$ indicates that the specific volume for a patient tends to increase with subsequent visits. Similarly, a negative value for $Z_{vol}$ indicated that the volume decreases over time.

\textbf{Grey Matter: } We observe that in the patients with volumetric data spanning four visits, $S= -6$, $Var(S) = 8.67$ and $Z_{vol} = -1.698$. As per the standard normal distribution table, the null hypothesis is rejected, and the $H_d$ alternate is accepted where the data follows a monotonic downward curve with a 95.54\% confidence.\newline Similarly, for the patients with three visits, the calculated metrics are $S= -3$, $Var(S) = 3.67$ and $Z_{vol} = -1.05$. Thus, the grey matter decreases monotonically with 85.31\% confidence for patients with three visits and 95.54\% with four visits.

\textbf{Cerebrospinal Fluid: } For a majority of patients with 3 visits (1,6,7,8,10,11,12,13), it is observed that the CSF metric monotonically increases with 85.31\% confidence after calculating the Mann Kendall parameters. As an outlier, only Patient 15 decreases monotonically with 85.31\% confidence. Patient 9 increases with a confidence of 69.85\% score.\newline For patients with 4 visits (3,4,5,14), all of them monotonically increase with confidence of 95.54\%. Patient 2, considering the dip in volume in the third visit, only increases overall with a confidence of 62.93\%.

\textbf{White Matter: } No majority trend is identified as all the patients have individual monotonic increases or decreases as per pathological diagnosis (reasons explained in next section).  Some patients show an increase in white matter volumes, like P3, P4 (63.31\%), P6, P7, P12, P13, all of which with 85.31\% confidence. Whereas patient 1 shows a decrease in volume with an 85.31\% confidence score, and patient 2  with a 95.45\% decrease.  Others also have intolerable dips or surges in middle visits, like Patients 8, 9, 10, 11 and 15, which cannot be substantiated as a majority trend.

\section{Discussion}
Due to limited access to computational resources, we have considered MR scans of twenty patients (with three or more visits) from the ADNI database and compared their volume changes across three different tissue classes, namely cerebrospinal fluid (CSF), white matter (WM) and grey matter (GM). Of the twenty patients, fifteen were chosen with a common pattern and their corresponding volumes were plotted across the three or four visits. The few inconsistencies left out can be attributed to several factors, such as responsiveness to symptom treatment, inaccurate MR scan acquisition, or incorrect segmentation due to spurious noise and other correlated abnormalities. The mean and standard deviation of each volume calculated and reported in Table~\ref{tab:gmv}, \ref{tab:csfv} and \ref{tab:wmv} shed light on the behaviour of volume changes and subsequent trend analysis substantiate it.  

In general, dementia is considered a grey matter degenerative disease as the common symptoms are congruent to a gradual loss of grey matter volume, which is associated with memory and emotions~\cite{GMpaper,GMUses}. Alzheimer's affected patients show a massive loss in grey matter volume when compared to normal control counterparts. Our experimented volumes are consistent with similar clinical findings as we observe an exhaustive general decline in GM volume in all the patients, as shown in Fig~\ref{fig:gmvolume} over the prognosis across several visits. According to our Mann-Kendall trend analysis, grey matter decreases across the visits with 95.54\% confidence with four visits and 85.31\% confidence with three visits. We can thus deduce that the decline in grey matter is heavily correlated with the disease. 

Similarly, when Fig~\ref{fig:csfvolume} and subsequent Mann Kendall trend analysis is observed, almost all of the patients have an increase steadily over all the visits, whereas some of them, like patients 2 and 9, have a negligible dip in the middle but are overall increasing (62.93\% and 69.85\% increasing respectively). Only patient 15 has shown a monotonic decrease (85.31\%) in CSF volume. It is thus imperative that an intolerable or impaired CSF shifting trending towards an increase in volume should also be considered as a common biomarker for the affliction, which is also consistent with clinical trials~\cite{CSF3,CSF,CSF2}. 

Fig~\ref{fig:wmvolume} shows the white matter volume changes in AD patients, and even though a majority trend could not be identified from the line plot and subsequent Mann-Kendall analysis where some increase monotonically and others decrease with random spikes of increase and decrease in the middle scans, certain specific abnormalities such as hyperintensities and lesion formation have been analyzed for their correlation with Alzheimer's in multiple studies~\cite{WM1,WM2,WM3}.

\vspace{-0.1cm}
\section{Conclusion}
Alzheimer's is one of the primary causes of mortality in the elderly population, which can not be treated by any existing preventative medication and leads to memory impairment and trauma for patients and caregivers alike. We have surveyed the possibility of imaging biomarkers pertaining to the disease and presented an easy-to-reproduce preprocessing pipeline with a wrapper script in R software. The pipeline includes common techniques such as inhomogeneity correction, skull stripping, image registration and intensity normalization and segmentation to determine the volume changes in three specific tissues (grey matter, white matter and cerebrospinal fluid) of AD-affected patients from structural MRI scans for longitudinal analysis. Experiments show that patients go through a gradual decline in grey matter volume and also exhibit increased intolerable shifting for cerebrospinal fluids, where the findings are substantiated by the Mann-Kendall trend analysis. Both inferences for GM and CSF volume change have existing clinical foundations for being relevant biomarkers for the disease, whereas hypertensive intensities and lesion formation in WM are studied for their involvement in AD. This work can be considered a primary step for future large-scale experimentation of the exhaustive MR scans of the ADNI dataset with more investigation on different thresholds and more accurate pre-processing and segmentation techniques. Instead of black-box automated analysis from raw MR scans or intensive annotation methods, this added step of calculating additional imaging features in the form of volumes interpreted from voxel changes related to the disease adds a layer of explainability with medical domain knowledge.
\section{Acknowledgements}
All the scans collected for analysis in preparation for this paper were accessed from the Laboratory of Neuro-imaging at the University of South California, which distributes ADNI data. ADNI is funded by the National Institute of Aging, the National Institute of Biomedical Imaging and Bioengineering, and through generous contributions from the following: AbbVie, Alzheimer's Association; Alzheimer's Drug Discovery Foundation; Araclon Biotech; BioClinica, Inc.; Biogen; Bristol-Myers Squibb Company; CereSpir, Inc.; Cogstate; Eisai Inc.; Elan Pharmaceuticals, Inc.; Eli Lilly and Company; EuroImmun; F. Hoffmann-La Roche Ltd and its affiliated company Genentech, Inc.; Fujirebio; GE Healthcare; IXICO Ltd.; Janssen Alzheimer Immunotherapy Research \& Development, LLC.; Johnson \& Johnson Pharmaceutical Research \& Development LLC.; Lumosity; Lundbeck; Merck \& Co., Inc.; Meso Scale Diagnostics, LLC.; NeuroRx Research; Neurotrack Technologies; Novartis Pharmaceuticals Corporation; Pfizer Inc.; Piramal Imaging; Servier; Takeda Pharmaceutical Company; and Transition Therapeutics. As such, the investigators within the ADNI contributed to the design and implementation of ADNI and/or provided data but did not participate in the analysis or writing of this report. A complete listing of ADNI investigators can be found at: \href{http://adni.loni.usc.edu/wp-content/uploads/how_to_apply/ADNI_Acknowledgement_List.pdf}{ADNI Acknowledgement}. The study also deeply benefits from the open-source course on Introduction to Neurohacking in R  offered by Johns Hopkins University.

\end{document}